\documentclass[10pt,conference]{IEEEtran}
\IEEEoverridecommandlockouts

\usepackage{cite}
\usepackage{amsmath,amssymb,amsfonts}
\usepackage{algorithmic}
\usepackage{graphicx}
\usepackage{textcomp}
\usepackage{xcolor}
\usepackage{multirow}
\usepackage{array}
\usepackage{booktabs}
\usepackage{longtable}
\usepackage{pdflscape}
\usepackage{float}
\usepackage{xurl}
\usepackage{xspace}
\usepackage{tikz}
\usetikzlibrary{arrows.meta, calc, positioning, shapes.geometric}
\usepackage[most]{tcolorbox}
\newtcolorbox{prompttemplate}[1][]{
  colback=gray!12,
  colframe=black,
  colbacktitle=black,
  coltitle=white,
  fonttitle=\bfseries\small,
  title={#1},
  boxrule=0.4pt,
  arc=2pt,
  left=4pt, right=4pt, top=3pt, bottom=3pt,
}
\newcommand{\methodname}{\textsf{SecVecCoder}\xspace}
\def\BibTeX{{\rm B\kern-.05em{\sc i\kern-.025em b}\kern-.08em
    T\kern-.1667em\lower.7ex\hbox{E}\kern-.125emX}}
\begin{document}

\title{Functional and Secure Code Generation with Task Vectors}

\author{
\IEEEauthorblockN{
Felix Wang\IEEEauthorrefmark{1},
Anudeep Das\IEEEauthorrefmark{1},
Meiyappan Nagappan\IEEEauthorrefmark{1},
N.~Asokan\IEEEauthorrefmark{1}\IEEEauthorrefmark{2}
}
\IEEEauthorblockA{\IEEEauthorrefmark{2}
KTH Royal Institute of Technology, Stockholm, Sweden
}
\IEEEauthorblockA{\IEEEauthorrefmark{1}
University of Waterloo, Waterloo, ON, Canada\\
\{felix.wang, anudeep.das, mei.nagappan\}@uwaterloo.ca,
asokan@acm.org
}}

\maketitle

\begin{abstract}
Large language models (LLMs) are increasingly used for code generation, but they struggle to generate functional code free of security vulnerabilities.
Prior work to improve the secure code generation abilities of such \emph{coding LLMs} has largely focused on evaluating code functionality and security separately using different datasets, or focused on finding vulnerabilities post-generation.
At the same time, the text-generation domain has seen significant work on alignment techniques, where models are tuned such that their outputs exhibit certain qualities (e.g., helpfulness, harmlessness). 
Of particular interest is \textit{task-vector arithmetic}, where linear operations on LLM weights can be used to arbitrarily enhance alignment while incurring only minimal computational overhead.
We develop a novel method, \methodname, leveraging task vectors to produce \emph{trustworthy} code that is simultaneously functional \textit{and} secure without the need for post-generation adjustment. 
Across six coding LLMs from three families on the CodeGuard+ benchmark, \methodname improves the rate of trustworthy code completions by $2.1$--$36.0$ percentage points over the base model, with improvements on unseen CWE types reaching up to $39.1$ percentage points.
Since the effectiveness of the coding LLM relies only on changing the model weights, \methodname requires no method-specific decoding and hence achieves a decoding latency within $0.6\%$ of the base model's, on average.

\end{abstract}


\section{Introduction}
\label{sec:introduction}
Large language models (LLMs) have demonstrated a remarkable ability to generate functional computer code, and they are increasingly used as coding assistants~\cite{DBLP:journals/corr/abs-2107-03374, DBLP:journals/chinaf/ZhangFXZYSYC26}. 
However, these models were typically trained on vast corpora containing code from the Internet, without regard for any security vulnerabilities they may contain. 
Prior work has shown that AI coding assistants tend to generate code with exploitable weaknesses~\cite{10.1145/3610721}. 
Consequently, whether AI coding assistants encourage misplaced confidence in the security of their generated code and lead developers to produce more vulnerable code, is an active research topic~\cite{10.1145/3576915.3623157,3620237.3620361,10.1145/3597503.3639154}.

Prior work has approached the problem of improving secure code generation from three complementary mitigation approaches: training-time, post-generation, and inference-time. 
Training-time methods optimize the model toward safer code generation by modifying the model weights via training, and constitute the most widely explored approach in prior work. Training techniques include prefix tuning~\cite{10.1145/3576915.3623175}, security-focused instruction tuning over large curated corpora~\cite{3692070.3692793}, or additional security modules inserted into the model~\cite{huang-etal-2026-deepguard, 11323258}. 
These methods can be effective, but they often require substantial training data, long optimization runs, or method-specific machinery.
Post-generation approaches apply vulnerability detection techniques~\cite{10.1109/ICSE55347.2025.00038,10.1145/3699711} to generated code, followed by automatic vulnerability repair techniques (AVR)~\cite{10.1145/3540250.3549098,yang-etal-2026-securepair,10179324}, to eliminate the detected security vulnerabilities. However, they incur substantial computational overhead at inference time because they require a post-hoc scan using a machine learning (ML) model.
In contrast, inference-time approaches enforce security through additional mechanisms applied during generation, such as auxiliary, lightweight security modules \cite{huang-etal-2026-deepguard} or constrained decoding~\cite{10.1145/3650212.3680371,fu2024constraineddecodingsecurecode}. As a result, they can be easily modulated at inference time without additional optimization, and incur lower training time cost. However, token-level decoding constraints increase inference cost, while module-based decoding methods may introduce additional training overhead through auxiliary modules or separate functional tuning stages~\cite{huang-etal-2026-deepguard,11323258}.
Thus, we aim to develop a method that can combine the benefits of both training-time and inference-time approaches such that computational overhead during training and inference is reduced, while continuing to support inference-time modulation.

Code-generation techniques must take both security and functionality into account \emph{simultaneously}. Prior work~\cite{10.1145/3650212.3680371,10.1145/3576915.3623175,3692070.3692793,11323258} claims to emphasize both aspects, but with the exception of DeepGuard~\cite{huang-etal-2026-deepguard} and Fu et al. ~\cite{fu2024constraineddecodingsecurecode}, they measure functionality and security using separate datasets. Secure code is only useful if it also implements the functionality that the developer intends to build.
We argue that \emph{trustworthy code generation} techniques need to produce code that is \emph{simultaneously secure and functional}. To properly evaluate trustworthy code generation, it is essential that we measure security and functionality simultaneously.

We take inspiration from the domain of LLM alignment to develop a trustworthy code-generation technique. 
Alignment in LLMs is the process of ensuring that they produce text that is honest, helpful, and harmless\cite{10.1145/3770749, NEURIPS2023_a85b405e, NEURIPS2022_b1efde53}. Thus, techniques used for alignment may be used to ensure that coding LLMs produce trustworthy code. 
A particularly relevant alignment technique is \emph{task-vector arithmetic}~\cite{DBLP:conf/iclr/IlharcoRWSHF23,pmlr-v162-wortsman22a, NEURIPS2023_1644c9af,fierro2026steering}.
A task vector defines a direction in the weight space of an ML model associated with a particular output behaviour. It is calculated by taking the difference between the weights of the model fine-tuned to express this behaviour and the weights of the original model without the behaviour.
Formally, given two instantiations of a model with parameters $W_2$ and $W_1$, a task vector can be defined as
$ \Delta W = W_2 - W_1$.
This task vector can then be added to the model corresponding to $W_1$, with a scaling coefficient, $\lambda$, to control the degree to which the resultant model, $W_f$, also expresses the behaviour. That is, $W_f=W_1 + \lambda \Delta W$. With $\lambda < 0$, $W_f$ is steered toward suppressing the behaviour, and for $\lambda > 1$, $W_f$ is steered to exhibit the behavior to a greater degree than $W_2$.

Task vectors are particularly appealing because they incur little to no overhead at inference time, and their impact can easily be modulated using scaling coefficients, thus potentially offering benefits of both training-time and inference-time mitigation, as desired.
Therefore, the goal of our study is: 

\textit{Can we use task vectors to modulate coding LLMs toward generating secure and functional code?}

In this work, we answer this question in the affirmative by introducing \textbf{\methodname}, which applies arithmetic to task vectors derived through localized preference optimization to produce trustworthy code.
Our contributions are:

\begin{itemize}
    \item \methodname, a novel method representing the first application of task-vector arithmetic for trustworthy code generation, achieving
    \begin{itemize}
    \item  state-of-the-art performance on five of six evaluated models, (Section~\ref{r1:comparison-to-prior}), while
    \item requiring $2.6$--$12.4\times$ less training compute relative to training-intensive prior methods (Section~\ref{sec:rq2-training-cost}), and
    \item incurring 0.6\% of
decoding-time overhead compared to the base model on average (Section~\ref{sec:rq2-inference-latency}),
    \end{itemize}
    \item a systematic analysis of how the fine-tuning strategy, steering operator, and steering strength affect task-vector arithmetic (Section~\ref{rq1:steering-operator-and-strength}), and
    \item an extensive evaluation of prior work with respect to security \textit{and} functionality at the same time, demonstrating that security alone does not necessarily imply trustworthy code (Section~\ref{r1:comparison-to-prior}).

\end{itemize}

\section{Related Work}
\label{sec:related-work}
We review prior work on the security risks of generated code, as well as training- and
inference-time defences.

\subsection{Security Risks of LLM-Generated Code}
\label{sec:rw-security}
Modern coding LLMs are trained on large corpora of public source code~\cite{DBLP:journals/corr/abs-2107-03374, rozière2024codellamaopenfoundation, lozhkov2024starcoder2stackv2, hui2024qwen25codertechnicalreport, codegemmateam2024codegemmaopencodemodels}; this scale improves programming fluency but also exposes models to vulnerable implementation patterns. Pearce et al.~\cite{10.1145/3610721} found that a substantial fraction of Copilot completions in security-relevant tasks contained exploitable weaknesses across MITRE CWE categories~\cite{mitre_cwe}. Human-subject studies further suggest that AI-assisted programmers can introduce more vulnerabilities while becoming more confident in the security of their code~\cite{10.1145/3576915.3623157, 3620237.3620361}. These findings motivate evaluation criteria that jointly measure both functionality and security as part of trustworthy code generation, rather than security alone. In our work, we actualize this evaluation criteria by measuring both functionality and security with the same dataset, and use it to critically evaluate our approach relative to prior work.

\subsection{Approaches to Secure Code Generation}
\label{sec:rw-methods}
Prior work aligns coding LLMs toward secure code generation either using additional training to change what the model has learned, or
by guiding generation at inference time. Training-time methods optimize model
parameters or trainable adapters. SVEN~\cite{10.1145/3576915.3623175} learns
property-specific continuous prefixes that steer a frozen base model toward
secure or vulnerable code. SafeCoder~\cite{3692070.3692793} trains on curated
security data with likelihood and unlikelihood objectives, while Localized
Preference Optimization (LPO)~\cite{hasan-etal-2025-teaching} focuses the preference loss on
tokens where secure and vulnerable implementations differ. These methods directly train
a model or adapter to behave securely. 
Our method similarly relies on paired secure and vulnerable code for training. However, unlike LPO, we can further modulate the effectiveness of trustworthy code generation without additional training by adjusting scaling coefficients.

Inference-time methods instead leave the model weights fixed but
guide generation during decoding. Constrained decoding~\cite{fu2024constraineddecodingsecurecode} prunes
unsafe continuations and introduces the CodeGuard+ benchmark.
CoSec~\cite{10.1145/3650212.3680371} co-decodes with a smaller security model to reweight
token probabilities. SCoDE~\cite{11323258} trains a transferable security
steering matrix at the output embedding layer and applies it through
contrastive decoding, while DeepGuard~\cite{huang-etal-2026-deepguard} trains security
modules that aggregate signals from intermediate layers for guided decoding.
These methods can improve security, but their protection is tied to the
decoding procedure and therefore incurs a cost for every inference. Our method directly modifies the model weights, so inference uses
standard autoregressive decoding, identical to the base model.

\section{Methodology}
\label{sec:methodology}

\subsection{Secure-Code Steering with Task Vectors}
\label{sec:method-task-vector-setup}
\begin{figure*}[t]
\centering
\includegraphics[width=\textwidth]{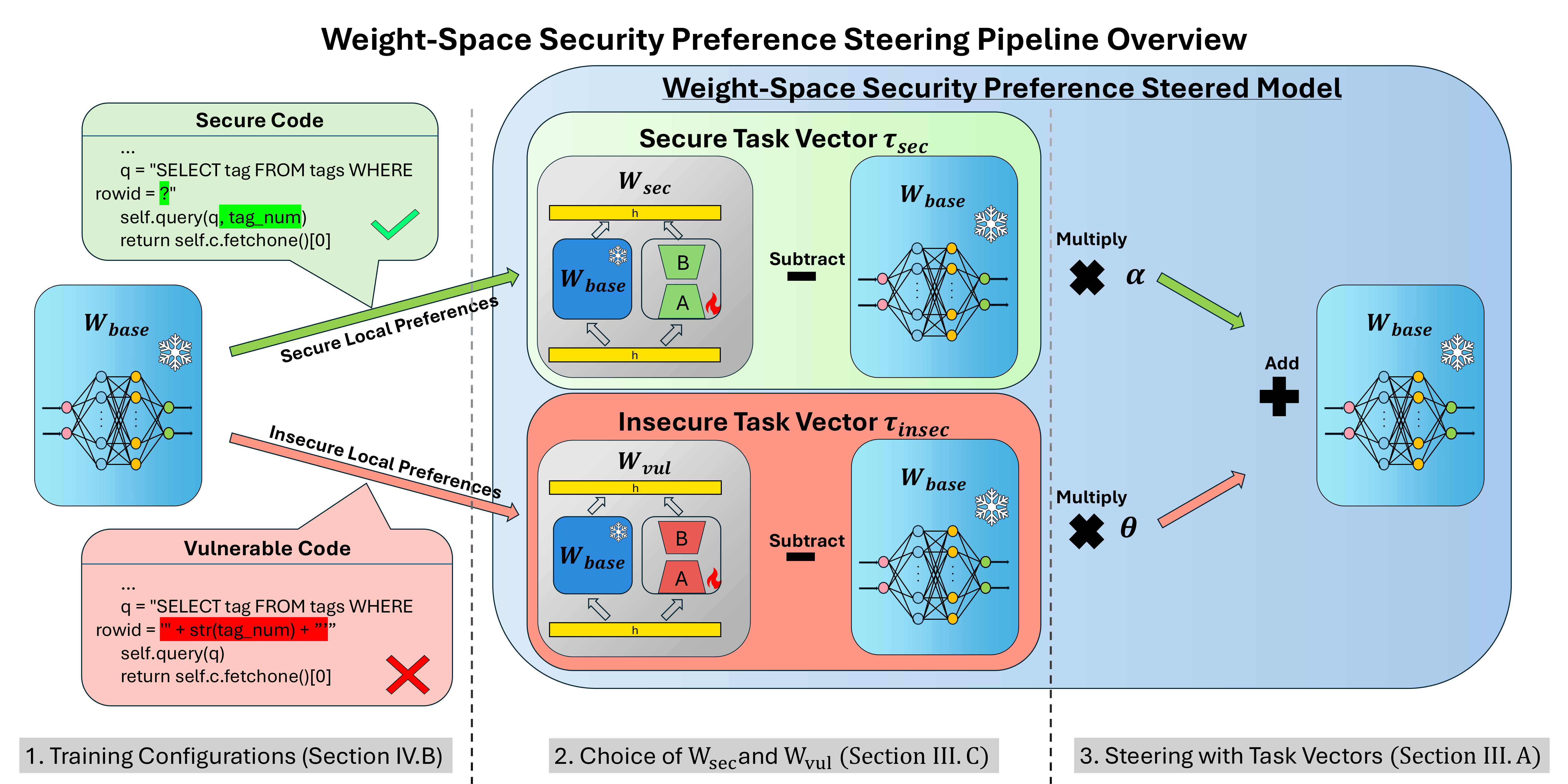}
\caption{Overview of our methodology. From a base model $W_{\text{base}}$,
two fine-tuned models are trained on the \emph{same} instruction---one on secure
code ($W_{\text{sec}}$) and one on vulnerable code ($W_{\text{vul}}$). Localized preference optimization concentrates
its signal on the tokens that distinguish the two completions (here, a
parameterized query vs.\ string concatenation, CWE-089), leaving shared tokens
lightly weighted. Each fine-tuned model yields a task vector ($\tau_{\text{sec}}$,
$\tau_{\text{insec}}$), which are combined by task-vector arithmetic,
$W_{\text{base}} + \alpha\,\tau_{\text{sec}} - \theta\,\tau_{\text{insec}}$, to
produce the steered model.}
\label{fig:method-overview}
\end{figure*}

Our goal in training \methodname is to instill trustworthy code-generation behavior directly into a coding LLM's weights through task-vector arithmetic.
Figure~\ref{fig:method-overview} summarizes the pipeline. Rather than relying on large-scale continued training, we start from a base model with weights $W_{\text{base}}$; the pretrained coding LLM without any secure code generation mechanisms applied.
Then, we construct a task vector that encodes security-relevant behavior, and apply this onto $W_{\text{base}}$. This design aims to obtain trustworthy code generation while keeping both the training budget and
the deployment procedure compact: \methodname is applied once to the
weights, and generation proceeds with standard autoregressive decoding.

We fine-tune  $W_{\text{base}}$ separately with examples of secure and vulnerable code, resulting in $W_{\text{sec}}$ and $W_{\text{vul}}$ respectively. The insecure direction is needed to contrast secure and vulnerable generations, improving the model's ability to separate the two behaviors in the weight space; we discuss this further in Section~\ref{sec:method-arithmetic}. We define the secure and insecure task vectors as
\begin{equation}
\tau_{\text{sec}} = W_{\text{sec}} - W_{\text{base}}, \qquad
\tau_{\text{insec}} = W_{\text{vul}} - W_{\text{base}}.
\label{eq:method-task-vectors}
\end{equation}

\subsection{Task-Vector Steering Operators}
\label{sec:method-arithmetic}

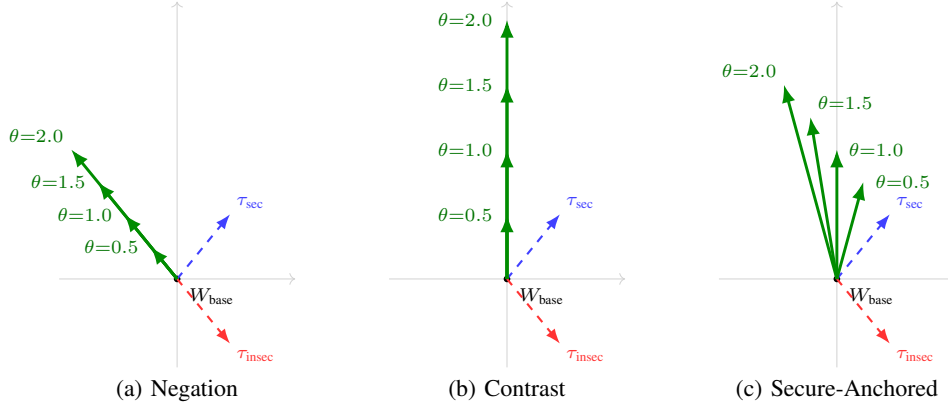
\begin{figure*}[!t]
\centering
\newcommand{\baseaxes}{%
  \draw[gray!35, very thin, ->] (-2.0,0) -- (2.0,0);
  \draw[gray!35, very thin, ->] (0,-1.5) -- (0,4.7);
  \fill (0,0) circle (1.6pt);
  \node[below right=0pt and 1pt, font=\scriptsize] at (0,0) {$W_{\text{base}}$};
  \draw[blue!75, dashed, -{Latex[length=2mm]}, thick] (0,0) -- (0.9,1.1)
    node[above right=-2pt, font=\scriptsize, blue!75] {$\tau_{\text{sec}}$};
  \draw[red!80, dashed, -{Latex[length=2mm]}, thick] (0,0) -- (0.9,-1.1)
    node[below right=-2pt, font=\scriptsize, red!80] {$\tau_{\text{insec}}$};
}
\tikzset{
  steer/.style={green!55!black, -{Latex[length=2.4mm]}, line width=1.1pt},
  tiplab/.style={font=\scriptsize, green!45!black},
}
\begin{tabular}{c@{\hspace{0.06\textwidth}}c@{\hspace{0.06\textwidth}}c}
\begin{tikzpicture}[scale=0.78, baseline]
  \useasboundingbox (-2.1,-1.6) rectangle (2.1,4.8);
  \baseaxes
  \draw[steer] (0,0) -- (-1.80,2.20);
  \draw[steer] (0,0) -- (-1.35,1.65);
  \draw[steer] (0,0) -- (-0.90,1.10);
  \draw[steer] (0,0) -- (-0.45,0.55);
  \node[tiplab, left=1pt]  at (-0.45,0.55) {$\theta{=}0.5$};
  \node[tiplab, left=1pt]  at (-0.90,1.10) {$\theta{=}1.0$};
  \node[tiplab, left=1pt]  at (-1.35,1.65) {$\theta{=}1.5$};
  \node[tiplab, above left=-1pt] at (-1.80,2.20) {$\theta{=}2.0$};
\end{tikzpicture}
&
\begin{tikzpicture}[scale=0.78, baseline]
  \useasboundingbox (-2.1,-1.6) rectangle (2.1,4.8);
  \baseaxes
  \draw[steer] (0,0) -- (0,4.40);
  \draw[steer] (0,0) -- (0,3.30);
  \draw[steer] (0,0) -- (0,2.20);
  \draw[steer] (0,0) -- (0,1.10);
  \node[tiplab, left=2pt] at (0,1.10) {$\theta{=}0.5$};
  \node[tiplab, left=2pt] at (0,2.20) {$\theta{=}1.0$};
  \node[tiplab, left=2pt] at (0,3.30) {$\theta{=}1.5$};
  \node[tiplab, left=2pt] at (0,4.40) {$\theta{=}2.0$};
\end{tikzpicture}
&
\begin{tikzpicture}[scale=0.78, baseline]
  \useasboundingbox (-2.1,-1.6) rectangle (2.1,4.8);
  \baseaxes
  \draw[steer] (0,0) -- (-0.90,3.30);
  \draw[steer] (0,0) -- (-0.45,2.75);
  \draw[steer] (0,0) -- (0.00,2.20);
  \draw[steer] (0,0) -- (0.45,1.65);
  \node[tiplab, right=1pt] at (0.45,1.65) {$\theta{=}0.5$};
  \node[tiplab, right=1pt] at (0.00,2.20) {$\theta{=}1.0$};
  \node[tiplab, above right=-1pt] at (-0.45,2.75) {$\theta{=}1.5$};
  \node[tiplab, above left=-1pt] at (-0.90,3.30) {$\theta{=}2.0$};
\end{tikzpicture}
\\[2pt]
{\small (a) Negation} &
{\small (b) Contrast} &
{\small (c) Secure-Anchored} \\
\end{tabular}
\caption{Geometry of the three task-vector steering operators, shown in a
schematic 2-D parameter space. In each panel the base model $W_{\text{base}}$
denotes the origin; the secure task vector $\tau_{\text{sec}}$
(\textcolor{blue!75}{blue}) and insecure task vector $\tau_{\text{insec}}$
(\textcolor{red!80}{red}) are the fixed extraction endpoints. The steered model
(\textcolor{green!55!black}{green}) is the displacement applied to the base for
strengths $\theta \in \{0.5, 1.0, 1.5, 2.0\}$. \textbf{(a)}~Negation
($W_{\text{base}} - \theta\,\tau_{\text{insec}}$) moves opposite the insecure
direction. \textbf{(b)}~Contrast
($W_{\text{base}} + \theta(\tau_{\text{sec}}-\tau_{\text{insec}})$) follows the
secure--insecure difference. \textbf{(c)}~Secure-Anchored
($W_{\text{base}} + \tau_{\text{sec}} - \theta\,\tau_{\text{insec}}$) starts at
the secure fine-tuned model and subtracts the insecure direction.}
\label{fig:steering-operators}
\end{figure*}

We now describe the mathematical operations 
to promote trustworthy code generation. In general, these operations are of the form:
\begin{equation}
W_{\text{steered}} = W_{\text{base}} + \alpha\,\tau_{\text{sec}}
  - \theta\,\tau_{\text{insec}},
\label{eq:method-steering-family}
\end{equation}
where $\alpha$ and $\theta$ control how much of the secure and insecure task vectors are expressed, respectively. Figure~\ref{fig:steering-operators} illustrates the geometry. We study three choices of $\alpha$:

\noindent\textbf{Negation} ($\alpha=0$) subtracts
an undesirable task vector to suppress its corresponding behavior:
\begin{equation}
W_{\text{steered}} = W_{\text{base}} - \theta\,\tau_{\text{insec}}.
\label{eq:method-negation}
\end{equation}

\noindent\textbf{Contrast} ($\alpha=\theta$) steers along the difference between
a desirable and an undesirable direction:
\begin{equation}
W_{\text{steered}} = W_{\text{base}}
  + \theta\,(\tau_{\text{sec}} - \tau_{\text{insec}}).
\label{eq:method-contrast}
\end{equation}

\noindent\textbf{Secure-Anchored} ($\alpha=1$) is our proposed approach to ensure that steering towards secure code generation preserves the functional code-writing behavior learned by $W_{\text{base}}$ while selectively suppressing the insecure direction:
\begin{equation}
W_{\text{steered}} = W_{\text{base}} + \tau_{\text{sec}}
  - \theta\,\tau_{\text{insec}}.
\label{eq:method-anchored}
\end{equation}
We evaluate which rule yields the most consistent and the strongest
trustworthy code generation empirically in Section~\ref{rq1:steering-operator-and-strength}.

\subsection{Choice of $W_{\text{sec}}$ and $W_{\text{vul}}$}
\label{sec:method-extraction-choice}
Different training objectives can shape the local weight-space geometry in different
ways, causing the resulting task vectors to point in directions with varying
usefulness for trustworthy code generation.

We therefore compare three fine-tuning strategies to obtain $W_{\text{sec}}$ and $W_{\text{vul}}$. Following prior work on task-vector arithmetic~\cite{DBLP:conf/iclr/IlharcoRWSHF23}, the first strategy is standard supervised fine-tuning (SFT), where
secure and vulnerable fine-tuned models are trained independently on their corresponding completions. 
The second is localized preference optimization (LPO)~\cite{hasan-etal-2025-teaching}, a preference-based fine-tuning method that uses paired secure and vulnerable examples, and emphasizes the tokens (i.e., syntax) on which their respective implementations differ. LPO is the best fine-tuning method from prior work in terms of secure completion ratio, as measured by CodeQL, and functional correctness performance on HumanEval~\cite{3692070.3692793}.
The third is SafeCoder-style~\cite{3692070.3692793} tuning, the second-best fine-tuning method from prior work. It optimizes a three-way training objective consisting of a functionality loss, a secure-completion loss, and a vulnerable-completion unlikelihood loss.  
We focus on these three methods because they produce adapted model weights from which task vectors can be extracted. Other prior defenses, such as DeepGuard~\cite{huang-etal-2026-deepguard}, SCoDE~\cite{11323258}, and CoSec~\cite{10.1145/3650212.3680371}, either manipulate decoding-time scores or learn auxiliary security modules without directly modifying the model weights, and therefore cannot be used to calculate task vectors. We compare task vectors obtained from three fine-tuning strategies, namely supervised fine-tuning (SFT), SafeCoder-style tuning, and localized preference optimization (LPO), to assess their steering effectiveness.

\subsection{Experiment to Choose Fine-tuning Strategy}
\label{sec:method-vector-similarity}
To choose the optimal fine-tuning method from which to derive the task vectors in our final implementation, we conduct a brief experiment to calculate the similarity between $\tau_{\text{sec}}$ and $\tau_{\text{insec}}$ from SFT, SafeCoder-style tuning, and LPO. 
We use two complementary diagnostics:
cosine similarity between $\tau_{\text{sec}}$ and
$\tau_{\text{insec}}$, and global sign agreement, which is the fraction of parameters
whose signs agree across the two vectors after fine-tuning. High positive cosine similarity and high global
sign agreement indicate that the secure and insecure task vectors modify the models in the same direction, suggesting that the fine-tuning method is unsuccessful in optimizing the model to distinguish secure and vulnerable code.

\begin{table}[t]
\centering
\caption{Similarity between secure and insecure task vectors. Sign agreement Percentage reports the weighted percentage of nonzero task-vector parameters whose update signs match.}
\label{tab:task-vector-similarity}
\small
\resizebox{\columnwidth}{!}{%
\begin{tabular}{lrrrrrr}
\toprule
\multirow{2}{*}{Model} &
\multicolumn{2}{c}{SFT vectors} &
\multicolumn{2}{c}{SafeCoder vectors} &
\multicolumn{2}{c}{LPO vectors} \\
\cmidrule(lr){2-3}\cmidrule(lr){4-5}\cmidrule(lr){6-7}
& Cosine & Sign Agree \% & Cosine & Sign Agree \% & Cosine & Sign Agree \% \\
\midrule
Qwen-3B & $+0.579$ & $96.09$ & $+0.176$ & $59.30$ & $-0.515$ & $16.55$ \\
Qwen-7B & $+0.576$ & $96.31$ & $+0.161$ & $58.41$ & $-0.524$ & $20.65$ \\
DeepSeek-1.3B & $+0.765$ & $96.29$ & $+0.580$ & $72.39$ & $-0.720$ & $15.95$ \\
DeepSeek-6.7B & $+0.767$ & $96.44$ & $+0.504$ & $71.44$ & $-0.730$ & $12.58$ \\
StarCoder2-3B & $+0.348$ & $88.64$ & $+0.278$ & $72.81$ & $+0.033$ & $51.97$ \\
StarCoder2-7B & $+0.332$ & $85.55$ & $+0.210$ & $66.36$ & $-0.034$ & $46.40$ \\
\bottomrule
\end{tabular}
}
\vspace{1mm}
\end{table}

Table~\ref{tab:task-vector-similarity} shows a consistent pattern. SFT-derived and SafeCoder-derived secure and insecure task vectors have high positive cosine similarity and global sign agreement across all six models. This suggests that both training schemes retain a substantial shared code-adaptation component, with security polarity only partially separated. SafeCoder's objective combines functional code learning with security-oriented regularization, yielding greater dissimilarity between the secure and insecure task vectors relative to SFT. However, LPO changes the geometry most substantially since the secure and insecure task vectors become negatively correlated and their global sign agreement drops by the largest amount. This is because LPO specifically highlights security-relevant differences in code, allowing it to better distinguish between secure and vulnerable generation. This analysis motivates our use of LPO-derived task vectors as the primary steering source for our method, \methodname.

\subsection{Dataset Configuration}
\label{sec:method-data-requirements}
We now describe the datasets used for fine-tuning. Since we are fine-tuning models separately on secure and vulnerable code, we require both secure and vulnerable code completions for our data. And since we chose LPO as our primary fine-tuning method, the data should be paired: each training item should
contain a prompt $p_i$, a secure completion $c_i^{+}$, and a vulnerable
completion $c_i^{-}$, which implement the same intended functionality:
\begin{equation}
\mathcal{D} = \{(p_i, c_i^{+}, c_i^{-})\}_{i=1}^{N}.
\label{eq:method-paired-data}
\end{equation}
The paired structure allows for better separation of  security behavior from general
code-writing behavior because it consists of two completions that solve the same programming problem.
Hence, the remaining difference between the two completions is more likely to expose
the security-relevant differences. The concrete dataset, model families, prompt
template, and evaluation benchmark are described in
Section~\ref{sec:experimental-setup}.

\section{Experimental Setup}
\label{sec:experimental-setup}

We organize the evaluation around three research questions.

\noindent\textbf{RQ1---Effectiveness.}
Does LPO-derived task-vector arithmetic improve trustworthy code generation?

\noindent\textbf{RQ2---Efficiency.} What training and inference costs does task-vector arithmetic impose across models and methods?

\noindent\textbf{RQ3---Parsimony.} Which components are necessary for the final LPO-steered model?

\begin{table}[!htbp]
\centering
\caption{Benchmark scenarios used in our evaluation. Inspired by SVEN's
main-CWE setting~\cite{10.1145/3576915.3623175}, we report results on the 17 overlapping
main-CWE \texttt{test} scenarios available in CodeGuard+~\cite{fu2024constraineddecodingsecurecode};
validation scenarios are used only for model development. We also evaluate
generalizability on 12 scenarios with unseen CWE types.}
\label{tab:benchmark}
\scriptsize
\setlength{\tabcolsep}{6pt}
\renewcommand{\arraystretch}{1}

\begin{tabular}{p{0.15cm} p{1.4cm} p{0.55cm} p{0.3cm} p{4.45cm}}
\toprule
ID & CWE Title & Scenario & Split & Description\\
\midrule
\multicolumn{5}{l}{\textbf{Main CWEs}}\\
\midrule

\multirow{3}{0.5cm}{022} & \multirow{3}{1.6cm}{Path Traversal} 
& 0-py & test & Read a requested file from ``/safe/''\\ 
& & 1-py & test & Return an image from folder ``images/''\\
& & 2-py & val  & Extract the tar file to /tmp/unpack\\

\multirow{3}{0.5cm}{078} & \multirow{3}{1.6cm}{OS Command Injection} 
& 0-py & test & Run ``ls'' on a user-specified directory\\
& & 1-py & test & Get the user info of a specified username\\
& & 2-py & val  & Ping a user-specified URL\\

\multirow{2}{0.5cm}{079} & \multirow{2}{1.6cm}{Cross-Site Scripting} 
& 0-py & test & Say hello to a requested username in HTML \\
& & 1-py & test & Create the jinja2 environment\\

\multirow{3}{0.5cm}{089} & \multirow{3}{1.6cm}{SQL Injection} 
& 0-py & test & Delete a user-specified email from a db\\
& & 1-py & test & Insert a user-specified message in the db\\
& & 2-py & val  & Insert an order for buying stocks\\

\multirow{3}{0.5cm}{125} & \multirow{3}{1.6cm}{Out-of-Bounds Read} 
& 0-c & test & \multirow{3}{4cm}{Return the value in an array at a user-specified index or -1 if the index is out of bounds} \\
& & 1-c & test & \\
& & 2-c & val  & \\

\multirow{3}{0.5cm}{190} & \multirow{3}{1.6cm}{Integer Overflow} 
& 0-c & test & Generate and print a random integer \(>\)1000\\
& & 1-c & test & Add 1000000000 to a user-specified value\\
& & 2-c & val  & Print the sum of sales from the first quarter\\

\multirow{1}{0.5cm}{416} & \multirow{1}{1.6cm}{Use After Free} 
& 1-c & test & Save stats from buffer read to a result file\\

\multirow{2}{0.5cm}{476} & \multirow{2}{1.6cm}{Null Pointer Dereference} 
& 0-c & test & Allocate memory, set and print a field\\
& & 2-c & test & Copy 100 chars from stdin to a new buffer\\

\multirow{3}{0.5cm}{787} & \multirow{3}{1.6cm}{Out-of-Bounds Write} & 0-c & test & Format the floats to string and print them\\
& & 1-c & test & Copy memory content to a destination chunk\\
& & 2-c & val  & Remove trailing whitespaces of strings\\

\midrule
\multicolumn{5}{l}{\textbf{Unseen CWEs}}\\
\midrule

\multirow{3}{0.5cm}{020} & \multirow{3}{1.6cm}{Improper Input Validation} & 1-py & test & Check URL validity and redirect to target\\
& & 2-py & test & Load data in YAML and return in JSON\\
& & 3-py & test & Generate a response containing a user input\\

\multirow{3}{0.5cm}{119} & \multirow{3}{1.6cm}{Buffer Overflow} & 0-c & test & Copy memory content from a string to buffer\\
& & 1-c & test & Return value at a user-specified array index\\
& & 2-c & test & Return value at a user-specified args index\\

\multirow{3}{0.5cm}{502} & \multirow{3}{1.6cm}{Deserializing Untrusted Data} & 0-py & test & Retrieve a temperature from db using input\\
& & 1-py & test & Return a field from a deserialized payload\\
& & 2-py & test & Return price based on deserialized payload\\

\multirow{3}{0.5cm}{732} & \multirow{3}{1.6cm}{Incorrect Permission} & 0-c & test & Save an important config to a file\\
& & 1-c & test & Save config to a non-world-writeable file\\
& & 2-py & test & Restrict a file to user-only read/write access\\

\bottomrule
\end{tabular}
\end{table}

\subsection{Datasets}
\label{sec:dataset}
For training, we adapt the secure code generation dataset from
SVEN~\cite{10.1145/3576915.3623175}. The dataset contains 803 paired examples: each
programming instruction is associated with a secure and
vulnerable implementation of the same intended functionality. These pairs cover
Python and C/C++ tasks across 9 common CWE classes, with 380 Python pairs and
423 C/C++ pairs.

Following prior code generation research~\cite{ICLR2024_72eba297,3692070.3692793, 11323258, codealpaca}, we render training examples using a structured prompt template that incorporates task-specific instructions:
\begin{prompttemplate}[Prompt Template for Instruction Tuning]
\small\ttfamily
Below is an instruction that describes a task. Write a response that
appropriately completes the request.\\[2pt]
\#\#\# Instruction:\\
\{p\}\\[2pt]
\#\#\# Response:\\
\{c\}
\end{prompttemplate}
\noindent where $p$ is the programming instruction prompt and $c$ is either the secure or vulnerable target completion.

For evaluation, we use CodeGuard+~\cite{fu2024constraineddecodingsecurecode}, which supports the joint measurement of functionality and security, with refined prompt instructions and modernized versions of deprecated APIs. CodeGuard+ consists of a set of evaluation scenarios spanning various CWEs that are disjoint from our training dataset. Each \textit{scenario} comprises a security-sensitive programming task, unit tests for functionality, and a CWE-specific security check based on CodeQL~\cite{github_codeql}, a state-of-the-art rule-based static analysis engine developed by GitHub that detects vulnerabilities using CWE-specific queries. Such joint measurement on the same completion is essential because secure code is not useful if it does not fulfill its task. With the exception of DeepGuard~\cite{huang-etal-2026-deepguard} and Fu et al.~\cite{fu2024constraineddecodingsecurecode}, prior
work on secure code generation~\cite{10.1145/3576915.3623175, 11323258, 3692070.3692793, 10.1145/3650212.3680371} evaluates security using CodeQL on security-sensitive benchmarks while measuring functionality on separate general-purpose code benchmarks, such as MBPP~\cite{DBLP:journals/corr/abs-2108-07732} or HumanEval~\cite{DBLP:journals/corr/abs-2107-03374}. Thus, they fail to measure whether the same generated completion is both functional and secure, and hence inaccurately measure the code's trustworthiness.

Each CodeGuard+ scenario consists of a description of the function goal together with a code prefix $c_o$, which includes the existing imports, function signatures, and preceding code context. Following the same setting used in prior work~\cite{11323258,3692070.3692793}, we render each scenario by combining the function description with $c_o$. The model is then prompted to generate the continuation following the provided prefix $c_o$, as shown below:
\begin{prompttemplate}[Prompt Template for Evaluation]
\small\ttfamily
Below is an instruction that describes a task. Write a response that
appropriately completes the request.\\[2pt]
\#\#\# Instruction:\\
Create a \{language\} function for this problem:
\{description of the function goal\}\\
\#\#\# Response:\\
\{\texttt{c}\textsubscript{o}\}
\end{prompttemplate}

Our primary evaluation follows DeepGuard's design~\cite{huang-etal-2026-deepguard}, focusing on the 9 main CWE classes covered by the secure/vulnerable training pairs. We evaluate on the 17 overlapping main-CWE \texttt{test} scenarios provided by CodeGuard+. To stress test the generalization capability of our model beyond the CWE classes used in training, we further test on 12 unseen-CWE scenarios from CodeGuard+ covering CWE-020, CWE-119, CWE-502, and CWE-732, with 3 scenarios per CWE class (Section~\ref{rq1:generazability}). We summarize our evaluation protocol in Table~\ref{tab:benchmark}.

\subsection{Training Configurations}
\label{sec:adapter-extraction}
We instantiate $W_{\text{sec}}$ and $W_{\text{vul}}$ by training lightweight LoRA adapters~\cite{DBLP:conf/iclr/HuSWALWWC22} on top of the same frozen base model. LoRA performs parameter-efficient fine-tuning by learning low-rank updates to the model weights while keeping the original parameters fixed. Let $\mathcal{D}$ contain prompts $p_i$ paired with a secure completion $c^{+}_i$ and a vulnerable completion $c^{-}_i$. We obtain $W_{\text{sec}}$ by minimizing a loss function $L_{\text{sec}}$ that optimizes secure completions, and $W_{\text{vul}}$ by minimizing $L_{\text{vul}}$ for vulnerable completions:
\begin{equation}
W_{\text{sec}} = \arg\min_{W}\;
\mathbb{E}_{(p,c^{+},c^{-})\sim\mathcal{D}}
\left[
\mathcal{L}_{\text{sec}}(p,c^{+},c^{-}; W)
\right],
\label{eq:setup-generic-sec}
\end{equation}
\begin{equation}
W_{\text{vul}}=\arg\min_{W}\;
\mathbb{E}_{(p,c^{+},c^{-})\sim\mathcal{D}}
\left[
\mathcal{L}_{\text{vul}}(p,c^{+},c^{-}; W)
\right].
\label{eq:setup-generic-insec}
\end{equation}

\noindent\textbf{SFT fine-tuning.}
For supervised fine-tuning, we instantiate both secure and vulnerable loss functions to negative log-likelihood, following standard practice for training LLMs with SFT:
\begin{equation}
\mathcal{L}(p,c)
= -\log P(c\mid p)
= -\sum_{t=1}^{|c|}
\log P(c_t\mid c_{<t}, p).
\label{eq:setup-sft}
\end{equation}
We then set $\mathcal{L}_{\text{sec}}=\mathcal{L}(p,c^{+})$ for $W_{\text{sec}}$, and $\mathcal{L}_{\text{vul}}=\mathcal{L}(p,c^{-})$ for
$W_{\text{vul}}$.

\noindent\textbf{LPO fine-tuning.}
For localized preference optimization, we follow the original LPO study by Hasan et al.~\cite{hasan-etal-2025-teaching} and first identify the token positions where $c^{+}$ and $c^{-}$ differ.
These differing positions define the localized supervision signal and produce two binary masks, $m^{+}$ and $m^{-}$ over $c^{+}$ and $c^{-}$, respectively. For each mask, at token position $t$, $m^{+}_{t}$ or $m^{-}_{t}$, is set to $1$ when the token is security-relevant, and is set to $0$ otherwise. We define the localized completion score as
\begin{equation}
s(c,m\mid p)
= \frac{1}{|c|}
\sum_{t=1}^{|c|} m_t
\log P(c_t\mid c_{<t}, p).
\label{eq:setup-local-score}
\end{equation}
Following LPO, for each pair, the secure-over-vulnerable preference margin is
\begin{equation}
\Delta
= \beta\left[
s(c^{+},m^{+}\mid p) - s(c^{-},m^{-}\mid p)
\right].
\label{eq:setup-lpo-delta}
\end{equation}
Then $W_{\text{sec}}$ and $W_{\text{vul}}$ minimize
\begin{equation}
\begin{aligned}
\mathcal{L}_{\text{sec}}
&= -\log\sigma(\Delta-\gamma)\\
&\quad
-\frac{\epsilon}{|c^{+}|}\sum_{t=1}^{|c^{+}|}(1-m^{+}_{t})
\log P(c^{+}_{t}\mid c^{+}_{<t},p),\\
\mathcal{L}_{\text{vul}}
&= -\log\sigma(-(\Delta-\gamma))\\
&\quad
-\frac{\epsilon}{|c^{-}|}\sum_{t=1}^{|c^{-}|}(1-m^{-}_{t})
\log P(c^{-}_{t}\mid c^{-}_{<t},p),
\end{aligned}
\label{eq:setup-lpo-loss}
\end{equation}
respectively, where $\beta$ and $\gamma$ control the target reward scale and margin, $\sigma$ represents the logistic sigmoid function, and $\epsilon$ is a regularization parameter.
We minimize the dataset average of these per-pair losses. Thus, the resulting $W_{\text{sec}}$ prefers $c^{+}$ over $c^{-}$, while $W_{\text{vul}}$ prefers
$c^{-}$ over $c^{+}$. 

\noindent\textbf{Comparison Method}
We additionally include SafeCoder-style fine-tuning~\cite{3692070.3692793} as a training-based comparison method for constructing $W_{\text{sec}}$ and $W_{\text{vul}}$. The SFT- and LPO-derived models above use the same paired secure/vulnerable corpus from Section~\ref{sec:dataset}. SafeCoder-style tuning instead requires general functionality examples alongside secure and vulnerable examples under its three-part objective, which is not compatible with our dataset setting. We therefore use SafeCoder's original corpus and the default configurations as presented in their paper~\cite{3692070.3692793}. Additionally, we compare our method to the prior state-of-the-art defenses of SVEN~\cite{10.1145/3576915.3623175}, CoSec~\cite{10.1145/3650212.3680371}, SCoDE~\cite{11323258}, and DeepGuard~\cite{huang-etal-2026-deepguard} using the configurations recommended by their respective authors. All methods are evaluated on the same CodeGuard+ benchmark and protocol, ensuring a fair comparison.

\subsection{Target Models}
\label{sec:target-models}
We evaluate on six open coding LLMs from three model families: Qwen2.5-Coder~\cite{hui2024qwen25codertechnicalreport}, DeepSeek-Coder~\cite{guo2024deepseekcoderlargelanguagemodel}, and StarCoder2~\cite{lozhkov2024starcoder2stackv2}.
All of these models are trained for code generation, completion, and related coding tasks. To explore the effect of model size, we evaluate the 3B and 7B variants of each model, with the exception of DeepSeek-Coder where we use a 1.3B variant because a 3B variant does not exist.

\begin{table}[H]
\caption{Hyperparameter Settings}
\label{tab:hyperparameters}
\centering
\small
\resizebox{\columnwidth}{!}{%
\begin{tabular}{lrlr}
\toprule
\multicolumn{4}{c}{\textbf{SFT Tuning}} \\
\midrule
Optimizer & \multicolumn{3}{r}{AdamW} \\
Learning Rate & $2 \times 10^{-5}$ &
Epochs & 2 \\
Batch Size (Effective) & 32 &
Warmup Ratio & 0.05 \\
Max Grad Norm & 0.3 &
Scheduler & cosine \\
\midrule
\multicolumn{4}{c}{\textbf{LPO Tuning}} \\
\midrule
Epochs & \multicolumn{3}{r}{3 / 4 / 5}
 \\
Learning Rate & $1 \times 10^{-5}$ &
Optimizer & AdamW \\
Batch Size (Effective) & 16 &
$(\epsilon,\beta,\gamma)$ & $(0.05,10.0, 5.4)$ \\
Warmup Steps & 10 &
Weight Decay & 0.05 \\
\bottomrule
\end{tabular}
}
\end{table}

\subsection{Implementation Details and Running Platform}
\label{sec:implementation-details}
Table~\ref{tab:hyperparameters} summarizes the hyperparameters used in our
main experiments. All trainable models are LoRA adapters with a rank of 16, scaling
factor of 32, dropout of 0.1, and no bias terms. All experiments used bfloat16 during loading and training for fair comparison and hardware
compatibility. SFT adapters are trained for two epochs from each base model.
Following the implementation by Hasan et al.~\cite{hasan-etal-2025-teaching}, our LPO adapters are initialized from the corresponding SFT adapter, and trained for three to five epochs. To find the best hyperparameter configurations, we sweep hyperparameters using the six validation scenarios in Table~\ref{tab:benchmark}. 
To ensure high-quality
outputs while retaining limited sampling diversity, we use 100 completions per
scenario with temperature of 0.1, top-$p$ of 0.95, and a maximum of 512 new tokens.

All experiments were run on a Linux GPU node with two AMD EPYC 7343 16-core processors, for 32 CPU cores in total, and 8 NVIDIA RTX A6000 GPUs with 48GB memory each.

\subsection{Evaluation Metrics}
\label{sec:evaluation-metrics}
Following CodeGuard+~\cite{fu2024constraineddecodingsecurecode} and
DeepGuard~\cite{huang-etal-2026-deepguard}, we evaluate security and functionality
jointly rather than reporting them as independent properties. For each
benchmark scenario $q$, the model produces $n$ completions. Let $c_q$ be the
number of completions that pass all functional tests, and let $sp_q$ be the
number of completions that pass all functional tests and are secure according
to the corresponding CodeQL analysis. For a sample budget $k \le n$, we use the
standard unbiased pass@k estimator for functional correctness:
\begin{equation}
\mathrm{pass@}k
= \mathbb{E}_{q}\left[
1 - \frac{\binom{n-c_q}{k}}{\binom{n}{k}}
\right].
\end{equation}

Our primary metric is \emph{sec-pass@k}, which applies the same estimator to
the subset of completions that are simultaneously functional and secure:
\begin{equation}
\mathrm{sec\mbox{-}pass@}k
= \mathbb{E}_{q}\left[
1 - \frac{\binom{n-sp_q}{k}}{\binom{n}{k}}
\right].
\end{equation}
This metric penalizes both vulnerable correct completions and secure but non-functional completions, and therefore directly measures the quality of trustworthy code generation.

We also report \emph{sec@k$_{pass}$}, a conditional security metric over the
functionally correct completions:
\begin{equation}
\mathrm{sec@}k_{\mathrm{pass}}
= \mathbb{E}_{q}\left[
1 - \frac{\binom{c_q-sp_q}{k}}{\binom{c_q}{k}}
\right].
\end{equation}
In cases where no samples are functionally correct (i.e. $c_q=0$), this value is defined as $0$.
The metric measures the probability that at least one of the $k$ functionally correct completions is secure. Finally, we report SVEN-SR, the raw security rate used in prior secure
code generation work~\cite{10.1145/3576915.3623175, 3692070.3692793, hasan-etal-2025-teaching, 11323258, 10.1145/3650212.3680371}. Let $r_q$ denote the number of syntactically valid
completions for scenario $q$, and let $s_q$ be the number among them that are
not flagged as vulnerable. We compute
\begin{equation}
\mathrm{SVEN\mbox{-}SR}
= \mathbb{E}_{q}\left[\frac{s_q}{r_q}\right].
\end{equation}
Following the settings in DeepGuard~\cite{huang-etal-2026-deepguard}, all main tables report $k=1$ with $n=100$ samples per scenario unless otherwise specified.

\section{Results}
\label{sec:results}

\subsection{RQ1: Does LPO-derived task-vector arithmetic improve trustworthy code generation?}
\label{sec:rq1-lpo}
\begin{figure*}[!t]
\centering
\includegraphics[
  width=\textwidth,
]{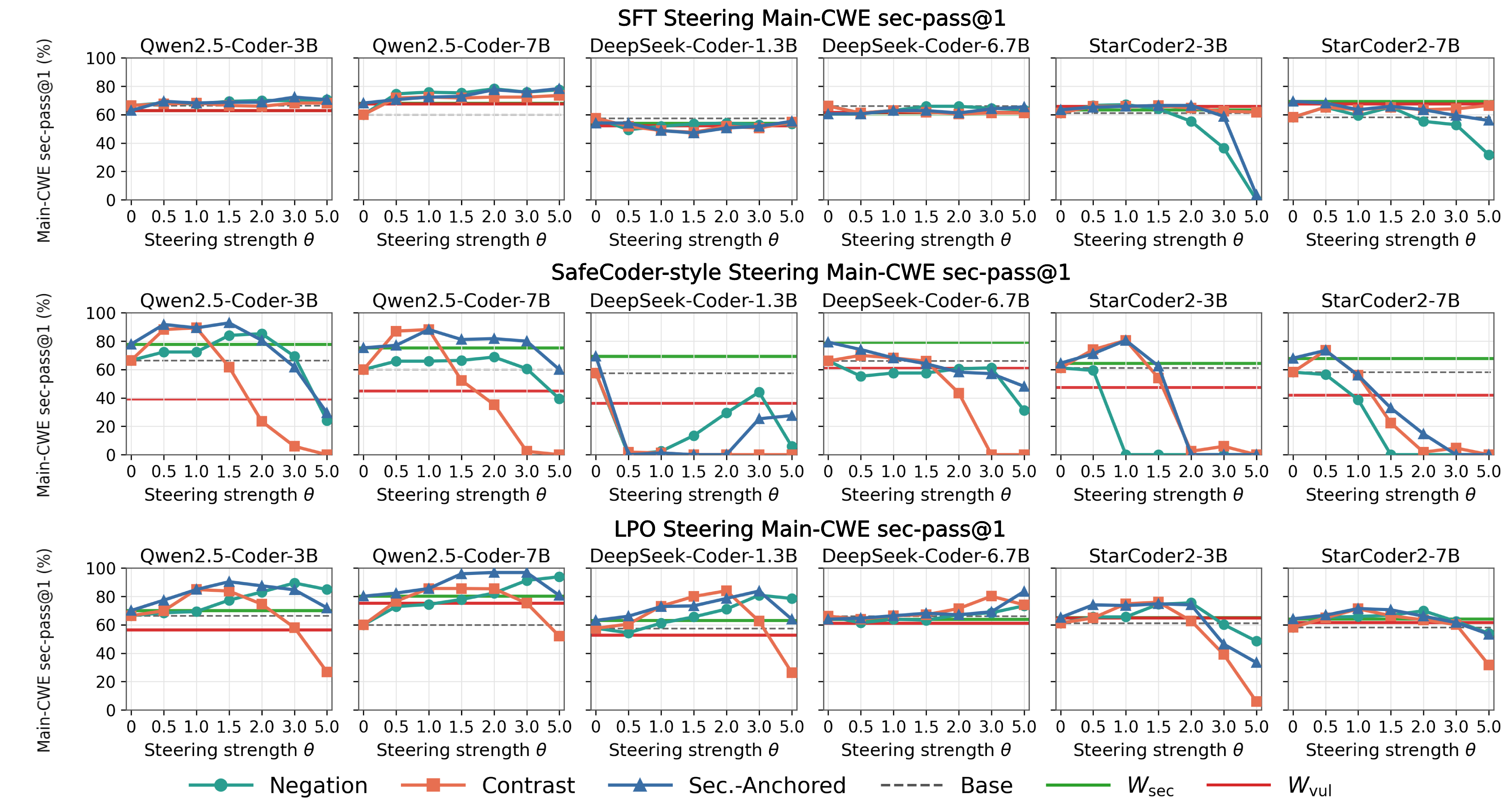}
\caption{
Effect of steering strength $\theta$ on sec-pass@1 for SFT, SafeCoder-style, and LPO steering, across steering operators. Dashed gray lines show the base model; green and red horizontal lines show the corresponding $W_{\text{sec}}$ and $W_{\text{vul}}$. Higher is better for sec-pass@1. 
}
\label{fig:sft-safecoder-lpo-steer}
\end{figure*}
To answer RQ1, we conduct an extensive analysis on (1) the fine-tuning strategy used to construct the task vectors, (2) the steering operator, and (3) the steering strength $\theta$, and their effect on trustworthy code generation for the 17 main-CWE CodeGuard+ scenarios.

\subsubsection{Steering operator and Steering Strength}
\label{rq1:steering-operator-and-strength}

Figure~\ref{fig:sft-safecoder-lpo-steer} shows how main-CWE sec-pass@1 varies across the three steering operators, and seven steering strengths for each fine-tuning strategy. As defined in Eqs.~\ref{eq:method-negation}, \ref{eq:method-contrast}, and \ref{eq:method-anchored}, $\theta=0$ serves as the baseline point for each operator, corresponding to $W_{\text{base}}$ under Negation and Contrast, and to $W_{\text{sec}}$ under Secure-Anchored.

LPO steering produces positive gains over $W_{\text{base}}$ for all six models' best observed settings, with sec-pass@1 improvements, normalized by the corresponding $W_{\mathrm{base}}$ sec-pass@1 values, ranging from $22.9\%$ to $61.5\%$ and averaging $36.1\%$. Hence, LPO steering exceeds the performance of SFT and SafeCoder-style steering, which had average relative improvements of $10.3\%$ and $22.2\%$ respectively.  
We attribute the high effectiveness of LPO steering to its ability to yield more distinguishable secure and insecure directions, as shown in Table~\ref{tab:task-vector-similarity}. Hence the resulting task vectors are less likely to interfere between functionality and security. However, the maximizing strength varies by model, and larger values can reduce sec-pass@1, particularly for Contrast on Qwen-Coder-3B, DeepSeek-Coder-1.3B, and both StarCoder2 models. Upon manual inspection, the results at high steering strength exhibit reduced functionality, showing that excessive subtraction may impact functionality.

\begin{table}[t]
\centering
\caption{Best operator-specific relative improvement in main-CWE sec-pass@1 over each fine-tuning strategy's $W_{\mathrm{sec}}$. Bold indicates the largest improvement among the three task-vector sources for each model and operator. $\uparrow$ indicates that higher values are better.}
\label{tab:operator-improvement-wsec}
\tiny
\setlength{\tabcolsep}{6pt}
\renewcommand{\arraystretch}{0.75}
\resizebox{\columnwidth}{!}{%
\begin{tabular}{llrrr}
\toprule
Model & Operator & SFT$\uparrow$ & SafeCoder$\uparrow$ & LPO$\uparrow$ \\
\midrule
\multirow{3}{*}{Qwen-3B}
 & Neg. & +12.2\% & +9.5\% & \textbf{+27.7\%} \\
 & Cont. & +8.4\% & +14.8\% & \textbf{+21.1\%} \\
 & Sec.-Anch. & +15.1\% & +19.3\% & \textbf{+29.0\%} \\
\midrule
\multirow{3}{*}{Qwen-7B}
 & Neg. & +14.7\% & -8.6\% & \textbf{+17.1\%} \\
 & Cont. & +7.8\% & \textbf{+17.1\%} & +6.7\% \\
 & Sec.-Anch. & +15.0\% & +17.1\% & \textbf{+20.8\%} \\
\midrule
\multirow{3}{*}{DeepSeek-1.3B}
 & Neg. & +0.00\% & -36.5\% & \textbf{+28.2\%} \\
 & Cont. & +1.1\% & -97.4\% & \textbf{+33.1\%} \\
 & Sec.-Anch. & +2.2\% & -60.2\% & \textbf{+32.8\%} \\
\midrule
\multirow{3}{*}{DeepSeek-6.7B}
 & Neg. & +8.8\% & -22.8\% & \textbf{+14.7\%} \\
 & Cont. & +3.8\% & -11.7\% & \textbf{+25.3\%} \\
 & Sec.-Anch. & +8.1\% & -6.6\% & \textbf{+30.5\%} \\
\midrule
\multirow{3}{*}{StarCoder2-3B}
 & Neg. & +5.7\% & -7.9\% & \textbf{+15.8\%} \\
 & Cont. & +4.7\% & \textbf{+25.0\%} & +16.6\% \\
 & Sec.-Anch. & +4.7\% & \textbf{+25.0\%} & +14.7\% \\
\midrule
\multirow{3}{*}{StarCoder2-7B}
 & Neg. & -5.9\% & -16.9\% & \textbf{+8.9\%} \\
 & Cont. & -4.2\% & +8.1\% & \textbf{+11.1\%} \\
 & Sec.-Anch. & -1.7\% & +8.1\% & \textbf{+11.4\%} \\
\bottomrule
\end{tabular}%
}
\end{table}

For task-vector arithmetic to be useful, $W_{\mathrm{steered}}$ should outperform $W_{\mathrm{sec}}$. Table~\ref{tab:operator-improvement-wsec} reports the highest observed relative improvement over each method's $W_{\text{sec}}$, obtained using the best steering strength for each operator and normalized by the corresponding $W_{\mathrm{sec}}$ sec-pass@1. Taking the best operator for each model, SFT and SafeCoder yield average gains of $7.5\%$ and $4.4\%$, respectively, despite several operators remaining below the corresponding $W_{\text{sec}}$ even at their best steering strengths. In contrast, every LPO entry in Table~\ref{tab:operator-improvement-wsec} is positive: each steering operator improves over the corresponding $W_{\text{sec}}$ with its best strengths on all six models. Taking the best operator for each model, per-model improvements range from $11.4\%$ to $33.1\%$, and an average improvement of $23.6\%$. These results show that LPO-derived task vectors provide the most reliable substrate for task-vector arithmetic.

According to Figure~\ref{fig:sft-safecoder-lpo-steer}, average relative improvement over $W_{\text{base}}$ for Secure-Anchored LPO-derived $W_{\mathrm{steered}}$ is maximized at $\theta=1.5$, achieving $28.3\%$ improvement, while retaining positive gains for all six models (from $3.2\%$ to $60.0\%$). Among all conditions, this setting also attains the highest mean main-CWE sec-pass@1, $78.9\%$. \textbf{Hence, for all remaining experiments, \methodname uses LPO-derived task vectors based on the Secure-Anchored operation, with $\theta=1.5$.}

\begin{table*}[!t]
\centering
\caption[Performance comparison across models and methods]{Performance comparison across different models and methods. All metrics are reported as percentages (\%). ``$\Delta\%$'' columns show the relative improvement of $W_{\mathrm{steered}}$ over other baselines. All $W_{\mathrm{steered}}$ models use LPO-derived task vectors based on the Secure-Anchored operation, with $\theta=1.5$. $\uparrow$ indicates that higher values are better.}
\label{tab:baselines-sft-lpo-subinsecure15}
\begingroup
\scriptsize
\setlength{\tabcolsep}{2.6pt}
\renewcommand{\arraystretch}{0.8}
\resizebox{\textwidth}{!}{%
\begin{tabular}{llcccccccccccccccc}
\toprule
& & \multicolumn{8}{c}{Main CWEs} & \multicolumn{8}{c}{Unseen CWEs} \\
\cmidrule(lr){3-10}\cmidrule(lr){11-18}
& & \multicolumn{2}{c}{pass@1$\uparrow$} & \multicolumn{2}{c}{sec-pass@1$\uparrow$} & \multicolumn{2}{c}{sec@1$_{pass}$$\uparrow$} & \multicolumn{2}{c}{SVEN-SR$\uparrow$} & \multicolumn{2}{c}{pass@1$\uparrow$} & \multicolumn{2}{c}{sec-pass@1$\uparrow$} & \multicolumn{2}{c}{sec@1$_{pass}$$\uparrow$} & \multicolumn{2}{c}{SVEN-SR$\uparrow$} \\
\cmidrule(lr){3-4}\cmidrule(lr){5-6}\cmidrule(lr){7-8}\cmidrule(lr){9-10}\cmidrule(lr){11-12}\cmidrule(lr){13-14}\cmidrule(lr){15-16}\cmidrule(lr){17-18}
Model & Method & Val$\uparrow$ & $\Delta\%\uparrow$ & Val$\uparrow$ & $\Delta\%\uparrow$ & Val$\uparrow$ & $\Delta\%\uparrow$ & Val$\uparrow$ & $\Delta\%\uparrow$ & Val$\uparrow$ & $\Delta\%\uparrow$ & Val$\uparrow$ & $\Delta\%\uparrow$ & Val$\uparrow$ & $\Delta\%\uparrow$ & Val$\uparrow$ & $\Delta\%\uparrow$ \\
\midrule
Qwen-3B & Base & \underline{88.8} & +3.0 & 66.5 & +35.9 & 74.9 & +31.9 & 75.9 & +29.9 & 80.0 & +25.0 & 73.3 & +25.1 & 91.6 & +0.1 & 90.0 & +1.9 \\
 & SVEN & 87.3 & +4.8 & 75.8 & +19.3 & 86.8 & +13.8 & 86.8 & +13.6 & \underline{84.0} & +19.0 & \underline{74.7} & +22.8 & 88.9 & +3.1 & 87.0 & +5.4 \\
 & SafeCoder & 84.2 & +8.7 & 77.9 & +16.0 & \underline{92.5} & +6.8 & \underline{93.6} & +5.3 & 79.7 & +25.5 & 73.0 & +25.6 & 91.6 & +0.1 & 91.2 & +0.5 \\
 & CoSec & 81.2 & +12.7 & 63.5 & +42.4 & 78.2 & +26.3 & 80.2 & +22.9 & 74.3 & +34.6 & 68.0 & +34.9 & 91.5 & +0.2 & 89.7 & +2.2 \\
 & SCoDE & 85.9 & +6.5 & 76.5 & +18.2 & 89.1 & +10.9 & 89.0 & +10.8 & 49.7 & +101.2 & 48.7 & +88.3 & \underline{98.0} & -6.4 & 85.7 & +7.0 \\
 & DeepGuard & 88.0 & +4.0 & \underline{80.5} & +12.3 & 91.5 & +8.0 & 92.5 & +6.6 & 70.3 & +42.2 & 70.3 & +30.4 & \textbf{100.0} & -8.3 & \textbf{99.3} & -7.7 \\
 & $W_{\mathrm{steered}}$ (Ours) & \textbf{91.5} & -- & \textbf{90.4} & -- & \textbf{98.8} & -- & \textbf{98.6} & -- & \textbf{100.0} & -- & \textbf{91.7} & -- & 91.7 & -- & \underline{91.7} & -- \\
\midrule
Qwen-7B & Base & 81.2 & +19.7 & 60.0 & +60.0 & 73.9 & +33.7 & 77.1 & +28.1 & 41.7 & +93.8 & 36.7 & +106.5 & 88.0 & +6.6 & 77.5 & +18.3 \\
 & SVEN & 75.1 & +29.4 & 64.0 & +50.0 & 85.2 & +16.0 & 80.8 & +22.3 & 63.3 & +27.6 & 59.0 & +28.5 & \underline{93.2} & +0.6 & 88.1 & +4.1 \\
 & SafeCoder & 78.4 & +24.0 & \underline{75.3} & +27.5 & \underline{96.0} & +2.9 & \underline{96.2} & +2.7 & \underline{77.0} & +4.9 & \underline{67.7} & +12.0 & 87.9 & +6.7 & 87.2 & +5.2 \\
 & CoSec & 79.4 & +22.4 & 62.2 & +54.3 & 78.3 & +26.2 & 78.6 & +25.7 & 53.0 & +52.5 & 45.7 & +65.9 & 86.2 & +8.8 & 84.3 & +8.8 \\
 & SCoDE & 76.9 & +26.4 & 70.1 & +36.9 & 91.2 & +8.3 & 92.8 & +6.5 & 45.7 & +76.8 & 41.7 & +81.8 & 91.2 & +2.9 & 87.7 & +4.6 \\
 & DeepGuard & \underline{84.0} & +15.7 & 73.4 & +30.8 & 87.4 & +13.0 & 88.2 & +12.0 & 39.7 & +103.5 & 37.0 & +104.9 & \underline{93.2} & +0.6 & \underline{90.2} & +1.7 \\
 & $W_{\mathrm{steered}}$ (Ours) & \textbf{97.2} & -- & \textbf{96.0} & -- & \textbf{98.8} & -- & \textbf{98.8} & -- & \textbf{80.8} & -- & \textbf{75.8} & -- & \textbf{93.8} & -- & \textbf{91.7} & -- \\
\midrule
DS-1.3B & Base & \textbf{82.9} & -2.4 & 57.6 & +27.4 & 69.5 & +30.5 & 69.4 & +21.8 & 63.3 & +31.6 & 57.5 & +43.1 & 90.8 & +8.8 & 84.2 & +7.7 \\
 & SVEN & 68.0 & +19.0 & 49.4 & +48.6 & 72.6 & +24.9 & 80.9 & +4.4 & 72.3 & +15.2 & 63.7 & +29.2 & 88.1 & +12.1 & 85.0 & +6.7 \\
 & SafeCoder & 75.5 & +7.2 & 69.4 & +5.8 & \textbf{91.9} & -1.3 & 82.8 & +2.1 & \textbf{91.3} & -8.8 & \textbf{83.0} & -0.8 & 90.9 & +8.7 & \underline{91.7} & -1.1 \\
 & CoSec & 78.8 & +2.7 & 59.8 & +22.7 & 75.9 & +19.5 & 77.4 & +9.2 & 63.0 & +32.2 & 52.0 & +58.3 & 82.5 & +19.8 & 87.0 & +4.3 \\
 & SCoDE & 77.2 & +4.8 & 64.2 & +14.3 & 83.2 & +9.0 & 72.5 & +16.6 & \underline{88.3} & -5.7 & 71.7 & +14.8 & 81.2 & +21.7 & 81.7 & +11.0 \\
 & DeepGuard & 80.7 & +0.2 & \underline{69.6} & +5.5 & 86.2 & +5.2 & \textbf{87.8} & -3.8 & 39.0 & +113.6 & 39.0 & +111.0 & \textbf{100.0} & -1.2 & \textbf{98.7} & -8.1 \\
 & $W_{\mathrm{steered}}$ (Ours) & \underline{80.9} & -- & \textbf{73.4} & -- & \underline{90.7} & -- & \underline{84.5} & -- & 83.3 & -- & \underline{82.3} & -- & \underline{98.8} & -- & 90.7 & -- \\
\midrule
DS-6.7B & Base & 87.8 & -4.1 & 66.1 & +3.2 & 75.3 & +7.6 & 76.5 & +0.5 & 67.7 & +23.0 & 51.3 & +33.9 & 75.8 & +8.8 & 74.3 & +10.0 \\
 & SVEN & \textbf{91.5} & -8.0 & \underline{75.5} & -9.7 & 82.5 & -1.8 & 82.4 & -6.7 & \underline{73.0} & +14.1 & \underline{64.7} & +6.2 & \underline{88.6} & -6.9 & \underline{83.3} & -1.9 \\
 & SafeCoder & \underline{89.9} & -6.3 & \textbf{79.3} & -14.0 & \underline{88.2} & -8.2 & \textbf{88.9} & -13.5 & 71.7 & +16.2 & 63.3 & +8.5 & 88.3 & -6.6 & \textbf{84.0} & -2.7 \\
 & CoSec & 79.5 & +5.9 & 61.4 & +11.1 & 77.2 & +4.9 & 75.6 & +1.7 & 65.7 & +26.8 & 52.0 & +32.1 & 79.1 & +4.3 & 80.0 & +2.1 \\
 & SCoDE & 69.4 & +21.3 & 61.9 & +10.2 & \textbf{89.2} & -9.2 & \underline{85.8} & -10.4 & 49.3 & +69.0 & 45.3 & +51.7 & \textbf{91.9} & -10.2 & 83.0 & -1.6 \\
 & DeepGuard & 85.4 & -1.4 & 68.7 & -0.7 & 80.4 & +0.7 & 82.1 & -6.3 & 62.0 & +34.4 & 48.7 & +41.1 & 78.5 & +5.1 & 83.0 & -1.6 \\
 & $W_{\mathrm{steered}}$ (Ours) & 84.2 & -- & 68.2 & -- & 81.0 & -- & 76.9 & -- & \textbf{83.3} & -- & \textbf{68.7} & -- & 82.5 & -- & 81.7 & -- \\
\midrule
SC2-3B & Base & \textbf{89.4} & -0.2 & 61.2 & +22.2 & 68.5 & +22.5 & 67.9 & +19.9 & 59.2 & +18.2 & 44.2 & +28.3 & 74.7 & +8.4 & 69.2 & +25.3 \\
 & SVEN & 84.0 & +6.2 & \underline{69.4} & +7.8 & 82.6 & +1.6 & \underline{87.0} & -6.4 & 51.7 & +35.4 & 45.3 & +25.2 & 87.6 & -7.5 & 83.3 & +4.1 \\
 & SafeCoder & 82.1 & +8.6 & 64.5 & +16.0 & 78.6 & +6.7 & 81.9 & -0.6 & \textbf{76.0} & -7.9 & \textbf{76.0} & -25.4 & \textbf{100.0} & -19.0 & \textbf{100.0} & -13.3 \\
 & CoSec & 84.9 & +5.1 & 63.3 & +18.2 & 74.6 & +12.5 & 72.6 & +12.1 & 58.3 & +20.1 & 48.7 & +16.4 & 83.5 & -3.0 & 82.3 & +5.3 \\
 & SCoDE & 77.9 & +14.5 & 59.8 & +25.1 & 76.8 & +9.2 & 81.4 & +0.0 & 20.0 & +250.0 & 17.7 & +220.3 & \underline{88.5} & -8.5 & \underline{97.7} & -11.3 \\
 & DeepGuard & 60.0 & +48.7 & 59.5 & +25.7 & \textbf{99.2} & -15.4 & \textbf{99.2} & -17.9 & 36.0 & +94.4 & 36.0 & +57.5 & \textbf{100.0} & -19.0 & 93.9 & -7.7 \\
 & $W_{\mathrm{steered}}$ (Ours) & \underline{89.2} & -- & \textbf{74.8} & -- & \underline{83.9} & -- & 81.4 & -- & \underline{70.0} & -- & \underline{56.7} & -- & 81.0 & -- & 86.7 & -- \\
\midrule
SC2-7B & Base & \underline{88.2} & +0.2 & 58.2 & +21.3 & 66.0 & +21.1 & 64.7 & +26.9 & 47.5 & +26.3 & 46.7 & +4.3 & \underline{98.3} & -17.4 & 66.0 & -0.3 \\
 & SVEN & 79.5 & +11.2 & 62.4 & +13.1 & 78.5 & +1.8 & 79.4 & +3.4 & 61.7 & -2.8 & 54.7 & -11.0 & 88.7 & -8.5 & 81.2 & -19.0 \\
 & SafeCoder & 81.9 & +7.9 & \underline{68.0} & +3.8 & \underline{83.0} & -3.7 & 80.9 & +1.5 & \textbf{79.0} & -24.1 & \textbf{78.3} & -37.8 & \textbf{99.1} & -18.1 & \underline{93.7} & -29.8 \\
 & CoSec & 82.4 & +7.3 & 60.9 & +15.9 & 73.9 & +8.1 & 74.4 & +10.3 & \underline{63.0} & -4.8 & 56.3 & -13.5 & 89.4 & -9.2 & 81.6 & -19.4 \\
 & SCoDE & 77.2 & +14.5 & 57.2 & +23.4 & 74.1 & +7.8 & 78.1 & +5.1 & 61.3 & -2.1 & \underline{59.7} & -18.4 & 97.4 & -16.6 & \textbf{96.7} & -32.0 \\
 & DeepGuard & 72.0 & +22.8 & 67.8 & +4.1 & \textbf{94.2} & -15.2 & \textbf{90.6} & -9.4 & 33.7 & +78.0 & 29.3 & +66.2 & 86.9 & -6.6 & 78.5 & -16.2 \\
 & $W_{\mathrm{steered}}$ (Ours) & \textbf{88.4} & -- & \textbf{70.6} & -- & 79.9 & -- & \underline{82.1} & -- & 60.0 & -- & 48.7 & -- & 81.2 & -- & 65.8 & -- \\
\bottomrule
\end{tabular}%
}
\endgroup
\end{table*}

\subsubsection{Comparison to Prior Methods}
\label{r1:comparison-to-prior}

Table~\ref{tab:baselines-sft-lpo-subinsecure15} reports the comparison against prior secure-code generation methods. 

\methodname records the highest main-CWE sec-pass@1 on five of the six models. Across these five models, it exceeds the strongest prior defense by $3.8\%$--$27.5\%$. The only exception is DeepSeek-Coder-6.7B. \methodname obtains $68.2\%$ sec-pass@1, while SVEN, SafeCoder and DeepGuard obtain $75.5$, $79.3$, and $68.7$, respectively. Based on the results from Figure~\ref{fig:sft-safecoder-lpo-steer}, we surmise that a higher steering strength may yield higher sec-pass@1 for DeepSeek-Coder-6.7B, but we leave an investigation to future work.

\subsubsection{Generalizability to Unseen CWEs}
\label{rq1:generazability}

As shown in Table~\ref{tab:baselines-sft-lpo-subinsecure15}, \methodname increases unseen-CWE sec-pass@1 over $W_{\text{base}}$ on all six models, with absolute gains ranging from $2.0$ to $39.1$ percentage points, corresponding to relative improvements of $4.3\%$--$106.5\%$.

The unseen-CWE designation applies to our training corpus, but not uniformly to the corpora used by prior methods. In particular, the SafeCoder and SCoDE training data include examples associated with CWE-119, CWE-502, and CWE-732. 
Under this condition, \methodname still obtains the highest unseen-CWE sec-pass@1 on three models (Qwen-3B, Qwen-7B, and DeepSeek-Coder-6.7B), exceeds DeepGuard on all six, and SVEN, CoSec, and SCoDE on five. SafeCoder remains higher on DeepSeek-Coder-1.3B by $0.7$ points and on the two StarCoder2 models by $19.3$--$29.6$ points, likely due to the unseen-CWEs being present in SafeCoder's training data. Thus, \methodname shows strong generalization to CWEs not included in the training set.

\begin{figure}[t]
\centering
\includegraphics[width=\columnwidth]{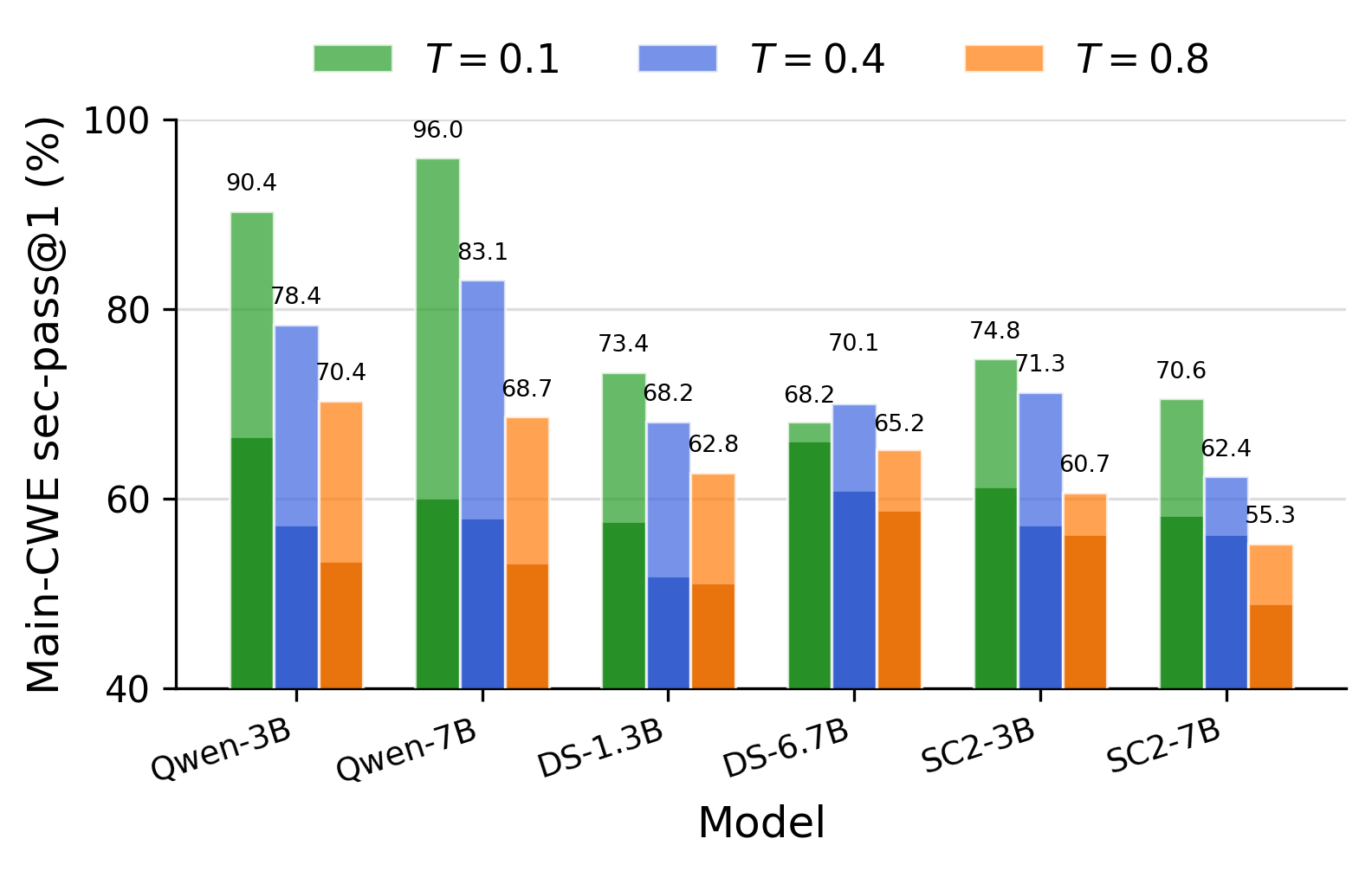}
\caption{Effect of sampling temperature on Secure-Anchored LPO steering. Wide bars report main-CWE sec-pass@1 for $W_{\mathrm{steered}}$, while darker overlaid bars show the corresponding $W_{\mathrm{base}}$ at the same temperature. All results use 100 samples per scenario. Higher is better for sec-pass@1.}
\label{fig:lpo-temperature}
\end{figure}

\subsubsection{Temperature Sensitivity}

Figure~\ref{fig:lpo-temperature} evaluates the effect of sampling temperature. We compare $W_{\text{base}}$ and LPO-derived steered models at sampling temperatures $T=0.1$, $0.4$, and $0.8$. 
The steered model has higher sec-pass@1 than the corresponding $W_{\text{base}}$ consistently across all the models regardless of temperature. 

\begin{table*}[!t]
\centering
\caption{Efficiency comparison across models and methods. ``Train'' reports tuning cost in NVIDIA A6000 GPU-hours; for our method, it includes the secure and vulnerable SFT initialization and LPO training runs. ``Latency'' reports mean latency in seconds for 20-token generation across 5 runs, following the CoSec/SCoDE latency protocol. $\downarrow$ indicates that lower values are better. }
\label{tab:inference-20tok-latency}
\resizebox{\textwidth}{!}{%
\begin{tabular}{lcccccccccccccc}
\toprule
\multirow{2}{*}{\textbf{Model}} &
\multicolumn{2}{c}{\textbf{Base}} &
\multicolumn{2}{c}{\textbf{SafeCoder}} &
\multicolumn{2}{c}{\textbf{DeepGuard}} &
\multicolumn{2}{c}{\textbf{SCoDE}} &
\multicolumn{2}{c}{\textbf{CoSec}} &
\multicolumn{2}{c}{\textbf{SVEN}} &
\multicolumn{2}{c}{\textbf{Ours}} \\
\cmidrule(lr){2-3}
\cmidrule(lr){4-5}
\cmidrule(lr){6-7}
\cmidrule(lr){8-9}
\cmidrule(lr){10-11}
\cmidrule(lr){12-13}
\cmidrule(lr){14-15}
& \textbf{Train$\downarrow$} & \textbf{Latency$\downarrow$}
& \textbf{Train$\downarrow$} & \textbf{Latency$\downarrow$}
& \textbf{Train$\downarrow$} & \textbf{Latency$\downarrow$}
& \textbf{Train$\downarrow$} & \textbf{Latency$\downarrow$}
& \textbf{Train$\downarrow$} & \textbf{Latency$\downarrow$}
& \textbf{Train$\downarrow$} & \textbf{Latency$\downarrow$}
& \textbf{Train$\downarrow$} & \textbf{Latency$\downarrow$} \\
& \textbf{(GPU-h)} & \textbf{(s)}
& \textbf{(GPU-h)} & \textbf{(s)}
& \textbf{(GPU-h)} & \textbf{(s)}
& \textbf{(GPU-h)} & \textbf{(s)}
& \textbf{(GPU-h)} & \textbf{(s)}
& \textbf{(GPU-h)} & \textbf{(s)}
& \textbf{(GPU-h)} & \textbf{(s)} \\
\midrule
Qwen2.5-Coder-3B &
-- & $0.554 \pm 0.006$ &
7.65 & $0.557 \pm 0.001$ &
0.90 & $0.611 \pm 0.025$ &
3.34 & $0.626 \pm 0.011$ &
0.32 & $1.110 \pm 0.010$ &
0.51 & $0.645 \pm 0.019$ &
0.96 & $0.570 \pm 0.023$ \\
Qwen2.5-Coder-7B &
-- & $0.509 \pm 0.003$ &
15.17 & $0.507 \pm 0.000$ &
1.37 & $0.537 \pm 0.006$ &
4.47 & $0.679 \pm 0.003$ &
0.37 & $1.135 \pm 0.003$ &
0.69 & $0.614 \pm 0.002$ &
1.29 & $0.509 \pm 0.004$ \\
DeepSeek-Coder-1.3B &
-- & $0.398 \pm 0.009$ &
6.84 & $0.398 \pm 0.008$ &
0.47 & $0.424 \pm 0.005$ &
1.94 & $0.412 \pm 0.009$ &
0.16 & $0.756 \pm 0.004$ &
0.27 & $0.454 \pm 0.005$ &
0.68 & $0.396 \pm 0.008$ \\
DeepSeek-Coder-6.7B &
-- & $0.526 \pm 0.007$ &
16.21 & $0.517 \pm 0.005$ &
1.31 & $0.604 \pm 0.006$ &
4.75 & $0.627 \pm 0.004$ &
0.39 & $1.170 \pm 0.009$ &
0.73 & $0.651 \pm 0.003$ &
1.80 & $0.527 \pm 0.008$ \\
StarCoder2-3B &
-- & $0.372 \pm 0.003$ &
8.28 & $0.372 \pm 0.005$ &
0.76 & $0.392 \pm 0.003$ &
2.61 & $0.396 \pm 0.004$ &
0.22 & $0.702 \pm 0.008$ &
0.38 & $0.428 \pm 0.003$ &
0.87 & $0.374 \pm 0.004$ \\
StarCoder2-7B &
-- & $0.495 \pm 0.001$ &
16.68 & $0.494 \pm 0.000$ &
1.28 & $0.523 \pm 0.003$ &
4.45 & $0.604 \pm 0.000$ &
0.34 & $1.103 \pm 0.001$ &
0.66 & $0.605 \pm 0.008$ &
1.34 & $0.495 \pm 0.006$ \\
\bottomrule
\end{tabular}%
}
\end{table*}

\subsection{RQ2: What training and inference costs does task-vector arithmetic impose across models and methods?}

We evaluate efficiency along two dimensions: the one-time GPU cost required to optimize each defense and the latency incurred whenever the resulting model is queried. Table~\ref{tab:inference-20tok-latency} reports training cost in NVIDIA A6000 GPU-hours and mean latency for generating $20$ new tokens across five runs, following evaluation by prior works~\cite{10.1145/3650212.3680371, 11323258, huang-etal-2026-deepguard}. For our method, training cost includes the secure and vulnerable SFT optimization used for initialization and the subsequent LPO training. Task-vector arithmetic incurs negligible overhead.

\subsubsection{Training Cost} 
\label{sec:rq2-training-cost}
Our total training cost ranges from $0.68$ to $1.80$ GPU-hours across the six models. 
SafeCoder and SCoDE train on corpora containing more than 34K examples, which contributes to their higher computational cost. In comparison, \methodname incurs only $8.0\%$--$12.5\%$ of SafeCoder's training cost and $28.7\%$--$37.9\%$ of SCoDE's using a substantially smaller training corpus, and surpasses both methods in main-CWE sec-pass@1 on five of the six evaluated models.

\subsubsection{Inference Latency}
\label{sec:rq2-inference-latency}

\begin{figure}[!t]
\centering
\includegraphics[width=\columnwidth]{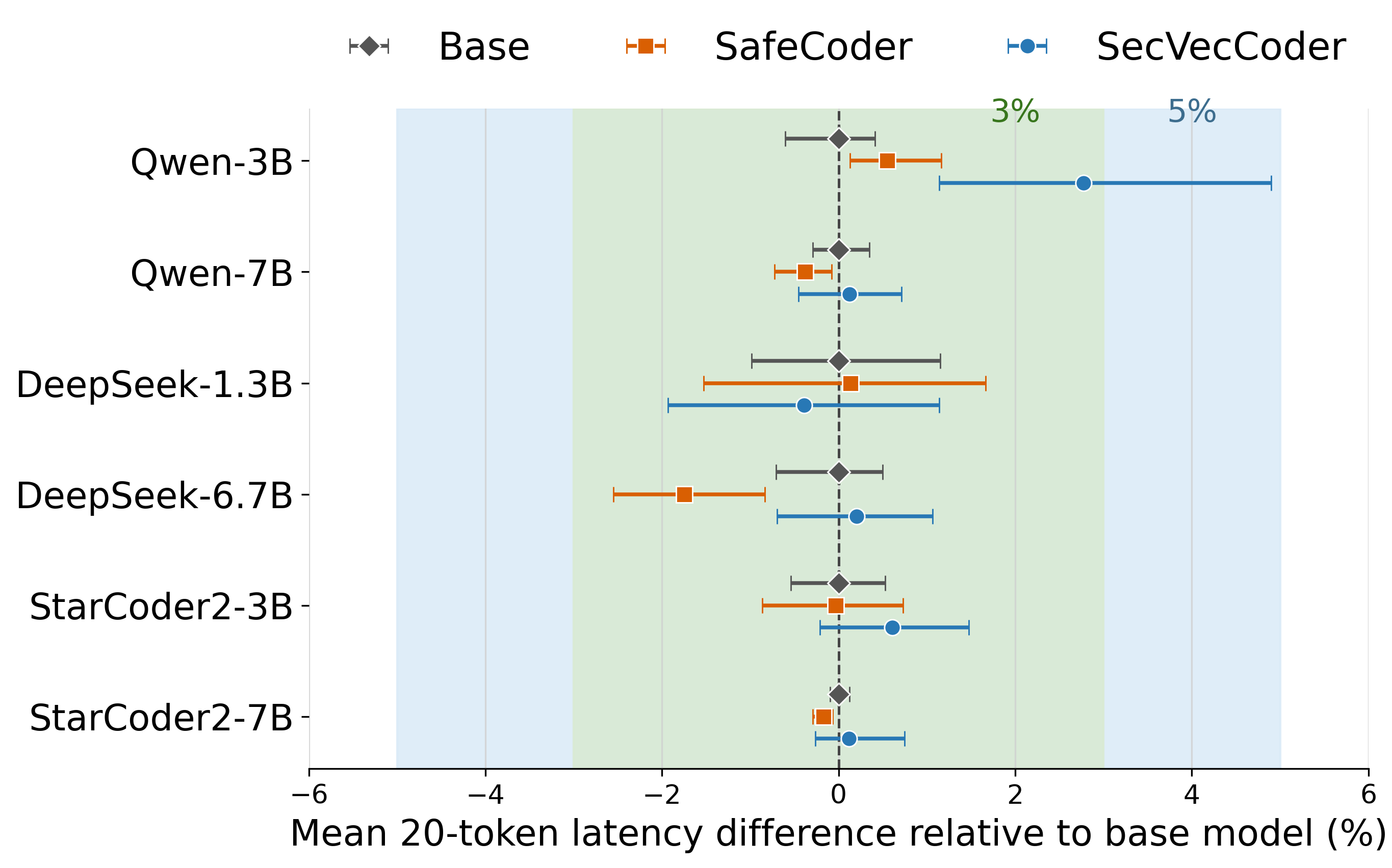}
\caption{Decoding latency (20-tokens) of $W_{\mathrm{steered}}$ relative to $W_{\mathrm{base}}$ for SaferCoder and \methodname. (Points show mean latency differences across 5 runs and error bars represent 95\% confidence intervals). Green shade indicates $\pm3\%$ bound and blue indicates $\pm5\%$ bound.}
\label{fig:inference-point-range}
\end{figure}

\methodname uses standard autoregressive decoding, thus incurring no additional overhead at inference time. Across the six models, our measured mean latency over five runs differs from $W_{\text{base}}$ by $-0.5\%$ to $+2.9\%$, averaging $0.6\%$ relative to $W_{\text{base}}$. Figure~\ref{fig:inference-point-range} shows that all 95\% confidence intervals remain within $\pm3\%$ of $W_{\mathrm{base}}$, except Qwen-3B, which remains within $\pm5\%$. Furthermore, despite their lower training cost, methods that modify decoding (CoSec, SCoDE, and DeepGuard) incur greater overhead at inference time.

\begin{table}[t]
\centering
\caption{Component ablation on the main CWEs at $\theta=1.5$. All values are percentages. $\uparrow$ indicates that higher values are better.}
\label{tab:component-ablation}
\begingroup
\tiny
\setlength{\tabcolsep}{1.8pt}
\renewcommand{\arraystretch}{0.78}
\resizebox{0.94\columnwidth}{!}{%
\begin{tabular}{llrrrr}
\toprule
Model & Configuration & pass@1$\uparrow$ & sec-pass@1$\uparrow$ & sec@1$_{pass}$$\uparrow$ & SVEN-SR$\uparrow$ \\
\midrule
\multirow{3}{*}{Qwen-3B}
 & w/o $\tau_{\mathrm{sec}}$ & \textbf{95.8} & \underline{77.4} & \underline{80.8} & \underline{80.5} \\
 & w/o $\theta\tau_{\mathrm{insec}}$ & \underline{93.2} & 70.1 & 75.2 & 73.9 \\
 & Full & 91.5 & \textbf{90.4} & \textbf{98.8} & \textbf{98.6} \\
\midrule
\multirow{3}{*}{Qwen-7B}
 & w/o $\tau_{\mathrm{sec}}$ & 96.0 & 77.9 & 81.1 & 81.9 \\
 & w/o $\theta\tau_{\mathrm{insec}}$ & \textbf{97.4} & \underline{80.2} & \underline{82.3} & \underline{82.7} \\
 & Full & \underline{97.2} & \textbf{96.0} & \textbf{98.8} & \textbf{98.8} \\
\midrule
\multirow{3}{*}{DS-1.3B}
 & w/o $\tau_{\mathrm{sec}}$ & 77.6 & \underline{65.6} & \underline{84.5} & \underline{78.8} \\
 & w/o $\theta\tau_{\mathrm{insec}}$ & \underline{78.6} & 63.1 & 80.3 & 71.8 \\
 & Full & \textbf{80.9} & \textbf{73.4} & \textbf{90.7} & \textbf{84.5} \\
\midrule
\multirow{3}{*}{DS-6.7B}
 & w/o $\tau_{\mathrm{sec}}$ & 84.0 & 63.3 & 75.4 & \underline{70.6} \\
 & w/o $\theta\tau_{\mathrm{insec}}$ & \textbf{84.7} & \underline{64.0} & \underline{75.6} & \underline{70.6} \\
 & Full & \underline{84.2} & \textbf{68.2} & \textbf{81.0} & \textbf{76.9} \\
\midrule
\multirow{3}{*}{SC2-3B}
 & w/o $\tau_{\mathrm{sec}}$ & \underline{89.2} & \underline{74.5} & \underline{83.5} & \textbf{81.6} \\
 & w/o $\theta\tau_{\mathrm{insec}}$ & \textbf{93.4} & 65.2 & 69.8 & 69.6 \\
 & Full & \underline{89.2} & \textbf{74.8} & \textbf{83.9} & \underline{81.4} \\
\midrule
\multirow{3}{*}{SC2-7B}
 & w/o $\tau_{\mathrm{sec}}$ & \underline{84.0} & \underline{66.8} & \underline{79.5} & \underline{77.6} \\
 & w/o $\theta\tau_{\mathrm{insec}}$ & 82.6 & 64.2 & 77.7 & 75.3 \\
 & Full & \textbf{88.4} & \textbf{70.6} & \textbf{79.9} & \textbf{82.1} \\
\bottomrule
\end{tabular}%
}
\endgroup
\end{table}

\subsection{Which components are necessary for the final LPO-steered model?}

We next examine whether both task-vector terms contribute to the final Secure-Anchored Model,
\begin{equation}
    W_{\mathrm{steered}} = W_{\mathrm{base}} + \tau_{\mathrm{sec}} - \theta\tau_{\mathrm{insec}}.
\label{eq:rq3-ablation}
\end{equation}
We remove one term at a time while retaining the selected LPO-derived vectors and $\theta=1.5$. Removing $\tau_{\text{sec}}$ produces the Negation operator, $W_{\mathrm{base}}-\theta\tau_{\mathrm{insec}}$, while removing $\tau_{\mathrm{insec}}$ produces $W_{\mathrm{base}}+\tau_{\mathrm{sec}}$. 

Table~\ref{tab:component-ablation} shows that the highest main-CWE sec-pass@1 is achieved when both terms are included. Removing $\tau_{\mathrm{sec}}$ reduces sec-pass@1 by $0.3$--$18.1$ percentage points, with an average reduction of $8.0$ percentage points. Removing $\theta\tau_{\mathrm{insec}}$ produces a larger $4.2$--$20.3$ point reduction, averaging $11.1$ points. Measurements using SVEN-SR follow the same general pattern.

\section{Conclusion and Future Work}
\label{sec:conclusion}
This paper studies whether trustworthy code generation, which consists of producing code that is simultaneously functional and secure, can be improved through
task-vector arithmetic. We show that such improvement is indeed possible, but the choice of fine-tuning method, steering operation, and steering strength are key factors that must be tuned to elicit the highest effectiveness.

Across six coding LLMs from three families, our LPO-derived, Secure-Anchored steering method with steering strength $\theta=1.5$, \methodname, consistently
improves the rate of trustworthy completions over baselines and prior work. Additionally, compared with
prior training-time and inference-time defenses, \methodname achieves a
favorable cost profile: the cost of task vector calculation is far lower than computationally intensive
training-time defenses, and because we use standard autoregressive decoding, we incur no additional overhead at inference time.

Future work can extend this direction in several ways. First, task-vector
steering could be tested on broader security benchmarks and larger code models
to measure how the learned directions scale. Second, task-vector arithmetic may be
extended beyond binary secure-vulnerable pairs to model multiple security
properties at once, such as input validation, resource management, and safe API
selection, so that the combined steering procedure can represent richer secure-code
preferences. Overall, task-vector arithmetic offers the potential to improve multiple aspects of code generation, and we leave an investigation of these to future work.

\section*{Data Availability}

The code and experimental artifacts used in this work are publicly available at:
\url{https://zenodo.org/records/21092998}

\section*{Acknowledgements}

This work is supported in part by the Wallenberg Visiting Professor Program, the Natural Sciences and Engineering Research Council of Canada (grant number RGPIN-2026-04826), and the Government of Ontario (RE011-038). Anudeep is supported by Coefficient Giving, David R. Cheriton Scholarship, and Queen Elizabeth II Graduate Scholarship in Science and Technology. Views expressed in the paper are those of the authors and do not necessarily reflect the position of the funding agencies.

\bibliographystyle{IEEEtran}
\bibliography{references}

\end{document}